# Isotropic Nanoscale Quantum Sensor at Room-Temperature

Daniel T. Möller[1], Baha Sakar[1], Ekrem T. Güldeste[1], Dhruv Wadhwani[1], Olt Gashi[1], Daniel A. García Vaca[1], Mokessh K. Ciwan[1], Simone P. Crozier[1], Christian Laube[2], Sirswa K. Shree Ram[1], Matteo Slaviero[1], Nabeel Aslam[1*]

[1]Felix-Bloch Institute for Solid State Physics, Leipzig University, Leipzig, 04103, Germany

[2]Leibniz Institute of Surface Engineering, Leipzig 04318, Germany

*Corresponding author: nabeel.aslam@uni-leipzig.de

**Abstract:**

**Color-center based quantum sensors provide nanoscale resolution under ambient conditions, yet their applicability remains limited. Because the quantization axes are locked to the host lattice, conventional color centers suffer severe signal loss in off-axis magnetic fields. To address this, we report an isotropic magnetometer enabled by the neutrally charged nitrogen-vacancy center ($NV^0$) in diamond. Here, spin-to-charge dynamics yield an $NV^0$-dark spin pair whose quantization axis dynamically aligns with the external field. Read out through NV charge-state-selective fluorescence, this system exhibits microsecond room-temperature coherence and nanotesla sensitivity for arbitrary field directions. We demonstrate alignment-free mapping of steep field gradients and single paramagnetic micro-targets, alongside isotropic readout from randomly oriented nanodiamonds. Resolving longstanding orientation constraints, this platform unlocks unrestricted nanoscale magnetometry across life sciences and quantum materials.**

**An isotropic spin-pair quantum sensor in diamond enables room-temperature nanoscale magnetometry across arbitrary fields.**

Quantum sensors (*1,2*) have transformed precision measurements across the physical and life sciences, yet existing tools, from superconducting quantum interference devices (SQUIDs) (*3*) to optically pumped atomic magnetometers (*4*), struggle to combine room-temperature operation, nanoscale resolution, and high-field dynamic range. Solid-state spin qubits - most prominently the negatively charged nitrogen-vacancy ($NV^-$) center in diamond (*5*) - have emerged as the premier platform to overcome these trade-offs (*6-9*). However, a fundamental bottleneck has constrained solid-state quantum magnetometers and their real-world applicability for decades: their spin quantization axes are rigidly fixed to the host crystal lattice. For off-axis field components beyond a few degrees from the quantization axis, crystalline anisotropy induces severe spin-state mixing (*10*), leading to sensitivity losses and eventually sensor breakdown. As a result, current solid-state sensors are often restricted to controlled laboratory settings and fail in key scenarios across different disciplines: they cannot map directional (in-plane versus out-of-plane) field responses *in-situ* in quantum materials, nor perform scanning-probe magnetometry over strongly magnetized ferromagnets due to spin mixing (*11*); they are severely restricted within steep unswitchable magnetic field gradients (*12*); and they suffer from significant signal and sensitivity losses caused by unpredictable rotational tumbling when deployed as nanodiamond probes in fluidic (*13*) or cellular environments. Overcoming these limitations requires an isotropic quantum sensor whose quantization axis aligns dynamically with the external magnetic field rather than the crystal lattice. While alternative solid-state systems - including spin-3/2 centers in silicon carbide (*14,15*) and spin-1/2 defects in 2D hexagonal boron nitride (*16-19*) - have been investigated, they inevitably trade off sensitivity or fail to maintain orientation independence regardless of magnetic field strength. Consequently, combining isotropic operation, nanoscale spatial resolution, biocompatibility, and room-temperature nanotesla sensitivity within a single solid-state platform has remained an unresolved challenge.

Here, we overcome this long-standing bottleneck by demonstrating an isotropic nanoscale quantum sensor in diamond operating under ambient conditions across arbitrary magnetic fields. We utilize an overlooked spin 3/2 manifold of the neutrally charged nitrogen-vacancy ($NV^0$) center in diamond, operating in its long-lived metastable state. Although the $NV^0$ charge state has been previously studied optically (*20-22*) and used to enhance the readout of the $NV^-$ spin (*23*), its spin degree of freedom in the metastable state has remained optically inaccessible until now, leading to its dismissal as an undesirable state (*20,24*). Here, we show that a spin-pair mechanism between the $NV^0$ center and an optically inactive dark spin enables both isotropic magnetic sensing and optical readout. First, the pair's central $|+1/2\rangle \leftrightarrow |-1/2\rangle$ transition provides a direction-independent magnetic response. Second, spin-to-charge conversion of the spin pair allows for optically detected magnetic resonance (SC-ODMR) via dual fluorescence emission bands (*25*). Room-temperature coherent control of the isotropic transition yields coherence times up to 3 μs, and multi-species detection alongside $NV^-$ centers is realized. We deploy the sensor to image steep magnetic field gradients where conventional $NV^-$ magnetometry fails and validate its orientation-independent response by magnetic imaging of paramagnetic microbeads under in-plane and out-of-plane bias fields. Finally, the readout of the spin pair system is demonstrated in individual nanodiamonds, establishing an alignment-free sensing platform ready for application in fluidic biological environments.

**Platform and isotropic readout mechanism**

For the realization of isotropic nanoscale magnetometry, we employ a microscopy setup with optical readout and microwave control. In contrast to the conventional $NV^-$ center, whose quantization axis is fixed to the diamond crystal lattice, the quantization axis of the $NV^0$-dark spin pair dynamically aligns with the external magnetic field (Fig. 1A). This scalar magnetometry across arbitrary field geometries is driven by a spin-to-charge (SC)-ODMR readout mechanism (Fig. 1B): Under 532 nm laser and microwave excitation, the system undergoes spin-selective charge dynamics which maps the state of the $NV^0$-dark spin pair onto the $NV^-/NV^0$ charge ratio (section S8). Spectroscopic measurements validate both this spin-pair model and the defect assignment (Fig. 1C). Under an on-axis field along the [111] lattice direction, continuous-wave (CW) SC-ODMR spectra reveal the three allowed transitions of the spin-3/2 manifold. The measured zero-field splitting (ZFS) ($2D \approx 3.37$ GHz) matches prior electron paramagnetic resonance (EPR) studies (*26*), identifying the spin 3/2 manifold in the metastable state of the $NV^0$ center. In addition, observed hyperfine splittings point to the involvement of nearby dark spins, for instance substitutional nitrogen (P1) centers.

To demonstrate the isotropic response, we measure the SC-ODMR signal of the central $|+1/2\rangle \leftrightarrow |-1/2\rangle$ transition while vertically displacing a permanent magnet by up to 25 mm (Fig. 1D). The resonance directly follows the field magnitude, enabling robust scalar mapping without any crystal alignment. Orientation independence is further validated by applying magnetic fields along orthogonal in-plane and out-of-plane axes (Fig. 1E). The frequency shift and signal contrast remain

invariant across all field orientations, circumventing the signal loss and complex vector reconstruction required by anisotropic defects like $NV^-$. The isotropic behavior establishes the $NV^0$-dark spin pair as a truly orientation-independent quantum sensor.

**Dual-band spin readout and multi-species magnetometry**

The distinct fluorescence emission spectra of the $NV^0$ and $NV^-$ centers enable simultaneous dual-band detection (Fig. 2A). Driving the central $|+1/2\rangle \leftrightarrow |-1/2\rangle$ transition produces anti-correlated signals between the two optical detection bands (Fig. 2B): a positive ODMR contrast in the $NV^-$ fluorescence channel accompanied by a negative contrast in the $NV^0$ channel. This complementary dual-band spin readout confirms that the signal arises from spin-dependent charge conversion between the two charge states (though the absolute sign of the contrast can vary across diamond samples, see section S3). In contrast to the central transition, the outer $|\pm1/2\rangle \leftrightarrow |\pm3/2\rangle$ transitions display an inverted contrast sign across both channels. This sign flip reflects a combination of the primary spin-pair charge dynamics and an additional readout contribution governed by spin-dependent metastable state lifetimes (section S6).

Extending this approach, we demonstrate multi-species magnetometry by tracking the $NV^-$ and $NV^0$ resonances as a function of applied magnetic field (Fig. 2C). Both species exhibit linear Zeeman scaling with a *g*-factor consistent with an electron spin, enabling concurrent vector ($NV^-$) and orientation-independent scalar ($NV^0$-dark spin pair) magnetometry from a single detection spot. To confirm the structural origin of the $NV^0$ spin in the metastable state, we deliberately misaligned the magnetic field relative to the crystallographic axes and tracked the zero-field splitting evolution (section S4). The observed angular dependence matches the characteristic $C_{3v}$ point-group symmetry of the $NV^-$ center, confirming that $NV^0$ in its metastable manifold retains the lattice symmetry predicted by density functional theory (*27,28*).

**Coherent control and metastable state dynamics**

To probe spin-dependent dynamics, we applied a resonant microwave pulse, followed by an optical readout of the spin state. The resulting fluorescence response reveal a stark contrast with the $NV^-$ center: while the $NV^-$ spin signal decays on a ~300 ns timescale set mainly by its singlet state lifetime, the signal of the $NV^0$-dark spin pair persists beyond 50 μs when probing the central $|+1/2\rangle \leftrightarrow |-1/2\rangle$ transition, and ~5 μs for the outer transitions (Fig. 3A). Separately, spin relaxation lifetime measurements yield decay times of ~110 μs for the central transition and 10–17 μs for the outer transitions (Fig. S10). The close agreement between these spin relaxation lifetimes and the decay of the optical contrast confirms that the observed SC-ODMR signal originates from the long-lived metastable manifold rather than ground-state dynamics.

These extended metastable state lifetimes enable coherent manipulation across the spin-3/2 manifold. We observe clear Rabi oscillations for the central $|+1/2\rangle \leftrightarrow |-1/2\rangle$ and the outer $|\pm1/2\rangle \leftrightarrow |\pm3/2\rangle$ transitions (Fig. 3C), with the Rabi frequency scaling linearly with the square root of microwave power, as expected for resonant quantum driving (Fig. 3D). Pulsed SC-ODMR spectroscopy further differentiates the hyperfine couplings of the spin pair (Fig. 3E and S1-S3): while the outer transitions exhibit hyperfine splittings matching the intrinsic nitrogen nuclear spin of $NV^0$ (*26*), the central transition displays signatures from P1 centers. Isotopic substitution ($^{14}N$ versus $^{15}N$) confirms that the central transition readout is mediated by the local dark-spin environment.

To quantify coherence properties, we perform Ramsey interferometry, Hahn echo, and XY8-2 dynamical decoupling sequences on the isotropic central transition at room temperature. These measurements yield a dephasing time $T_2^*$ of 99 ns and coherence times $T_2$ extending up to 3 μs (Fig. 3F). Demonstrating microsecond coherence confirms the practical viability of the $NV^0$- based platform as a quantum sensor under ambient conditions.

**Nanoscale magnetic imaging and orientation-independent sensing**

We demonstrate the practical utility of the isotropic $NV^0$-based sensor by measuring strong, spatially varying magnetic fields generated by a permanent magnet. CW-ODMR spectra recorded at different detection spots within the diamond illustrate this operational resilience (Fig. 4A). At a first detection spot (~1760 G), the local magnetic field vector is oriented such that both $NV^-$- and $NV^0$-based sensor yield resonance signals. Moving the optical detection spot to a second location changes both the magnetic field magnitude (~2110 G) and its angle relative to the crystal lattice. At this position, the increased off-axis field component induces severe spin-state mixing in $NV^-$, causing its ODMR contrast to vanish entirely. In contrast, the $NV^0$-dark spin pair's central transition, maintains a stable SC-ODMR signal, demonstrating immunity to the changing field vector direction. To extend these single-point measurements to spatial magnetic imaging, we mapped a 100 µm × 100 µm area of the diamond region (Fig. 4B). Across the entire field of view, the $NV^0$-based sensor produces a field distribution of the local magnetic field strength, whereas the $NV^-$ signal is completely quenched.

To validate the isotropic capability in biological proxies, we detect individual paramagnetic microbeads (~10 µm nominal diameter) drop-casted onto a diamond membrane (Fig. 4C), which serve as a model for cellular metal accumulation linked to neurodegenerative conditions (*29*). By applying bias fields along arbitrary orientations, we mapped the local magnetic fields *in-situ* without any sensor reconfiguration: an out-of-plane field yields a symmetric Gaussian profile (Fig. 4E), while in-plane biases along orthogonal x- and y-axes reveal characteristic dipole signatures (Fig. 4F and G).

Crucially, we extend the isotropic spin readout to individual nanodiamonds (nominal size ~ 250 nm) (Fig. 5). At an out-of-plane bias field of 2865 G we record CW SC-ODMR spectra from multiple randomly oriented single nanodiamonds, all displaying identical resonance frequencies. While $NV^-$ centers suffer from severe signal quenching and a wide spread of resonance frequencies due to varying crystallographic projections relative to the bias field, the $NV^0$-based central transition frequency and its linewidth remain strictly orientation-invariant. This demonstrates that $NV^0$-bearing nanodiamonds provide a robust, alignment-free platform for quantitative nanoscale magnetometry in randomly oriented environments.

**Discussion and Outlook**

The realization of the $NV^0$-based platform fundamentally changes magnetic field sensing with solid-state spin qubits. Rather than being limited by crystallographic orientation constraints, the $NV^0$-dark spin system adopts a quantization axis determined by the local field vector. As an optically addressable transducer, $NV^0$ translates spin-dependent charge exchange into dual-band fluorescence. Furthermore, while combining $NV^-$ and $NV^0$ yields a multi-modal sensing node, $NV^0$ alone can deliver complete, dual-mode magnetometry: its dark-spin-mediated central transition provides scalar readout, while its outer transitions, governed by orientation-dependent zero-field splitting, enable full 3D vector field reconstruction.

The dark spin mediated readout is remarkably robust across diverse diamond synthesis methods (CVD and HPHT), isotopic purity ($^{12}C$ vs. natural abundance) and nitrogen dopants ($^{14}N$ vs. $^{15}N$) (section S3). These findings confirm that SC-ODMR stems from intrinsic charge exchange dynamics rather than localized material artifacts.

We propose an initial theoretical framework for the SC-ODMR mechanism (see also section S8): Optical ionization from $NV^-$ initializes the metastable spin 3/2 $NV^0$ manifold, while the ejected electron is captured by a proximal defect to form a dark spin ($S$=1/2). Subsequent charge recombination is governed by the total spin angular momentum ($J'$) of the resulting $NV^0$-dark spin pair (Fig. 1B). Recombination is allowed only for joint states with $J'$=1 character, which matches the $S$=1 ground manifold of $NV^-$, whereas recombination from $J'$=2 character is forbidden. Consequently, individual joint states of the $NV^0$-dark spin pair show varying recombination rates. Resonant microwave driving on either or both spins transfers population between these $J'$ manifolds, modulating the charge recombination probabilities. Crucially, because these spin selection rules are rotationally invariant, the central $|-1/2\rangle \leftrightarrow |+1/2\rangle$ transition remains isotropic under arbitrary magnetic field orientations.

Beyond sensing, the coupled $NV^0$–dark spin pairs provide a versatile model system for investigating fundamental spin-dependent charge transport and exploring charge-mediated interactions for quantum information processing (*30*), a direction strongly supported by demonstrations of optical activation and charge transport between individual diamond color centers (*31*).

Presently, the $NV^0$-based central transition achieves a magnetic field sensitivity of ~70 nT/√Hz, within a ~390 μm³ collection volume (numerical aperture 1.49) at ~2% optical contrast (Fig. S9). Key optimization pathways include: Implementing solid-immersion lenses to boost collection efficiency, applying lock-in detection, and tailoring defect density to narrow the resonance linewidths. Additionally, using a multi-wavelength approach, one wavelength optimized for efficient $NV^0$ initialization and another for maximizing photon collection and charge stability, could significantly improve both photon counts and signal contrast. Together with dual-channel detection, these optimizations realistically project the sensitivity into the 100 pT/√Hz regime, while retaining isotropic operation (see also section S13).

The operational flexibility unlocks applications previously inaccessible to anisotropic defects. In bio-sensing, maintaining stable optical contrast under arbitrary bias fields enables real-time magnetic particle detection and continuous 3D tracking of functionalized microbeads, magnetic nanoparticle assemblies, and targeted carriers in complex media (*32*). Furthermore, $NV^0$-bearing nanodiamonds eliminate the directional blind spots and rotational signal loss that afflict $NV^-$ centers in dynamic environments. With the demonstrated dynamical decoupling yielding coherence times of $T_2$ = 3 μs, orientation-independent AC magnetometry and nano- and microscale NMR spectroscopy (*33,34*) becomes feasible, even with nanodiamonds as *in-situ* sensors. This capability opens pathways for real-time monitoring of hyperpolarized probes, such as [1-$^{13}$C]-pyruvate, extending macroscopic tracking of glycolytic flux and tumor metabolic heterogeneity (*35*) down to the single-cell level. Since the $NV^0$-dark spin system retains full addressability under arbitrary magnetic field angles, where $NV^-$ depolarizes (*36,37*), the system can serve as a robust source for hyperpolarization itself (*38*). Consequently, this can enable efficient polarization transfer from nanodiamonds (*39*) to surrounding nuclear spins, leveraging their high surface-to-volume ratio for enhanced hyperpolarization of external targets and their use as contrast agents.

Sub-nanometer spatial encoding for room-temperature magnetic resonance imaging (nanoMRI) (*40*) relies on applying steep magnetic field gradients. With $NV^-$ this is only possible with active switching and/or operating at relatively low fields (*41*), but for the platform introduced here no such limitations exist. Furthermore, operating reliably in high field gradients also facilitates the study of ferromagnetic domain structures (*42*) and phase transitions without signal quenching, while also enabling direct, alignment-free probing of nanoscale magnetic anisotropy. This capability is critical for characterizing quantum materials, including high-temperature superconductors (*43*), altermagnets (*44*), and topological spin textures (*45*). Extending this architecture to single $NV^0$-dark spin pairs in scanning-probe geometries (*46*) will furthermore enable high-resolution spatial mapping of magnetic structures without orientation constraints. In parallel, implementing photoelectrical readout schemes (*47*) will bypass optical photon-collection limitations, paving the way for fully integrated, chip-scale quantum sensors.

The $NV^0$-based system offers a direct solution to the orientation constraints of solid-state quantum sensors. By unlocking a previously inaccessible spin 3/2 manifold through dark-spin mediated charge dynamics, this work establishes a robust foundation for unrestricted quantum sensing across complex, arbitrarily oriented magnetic environments.

**Acknowledgments:** We thank Adam Gali, Jean-Philippe Tetienne, and Tim Fabian Prentke for helpful discussions, and Mohamed Abobeih for valuable feedback on the manuscript. We thank Jan Griebel for providing analytical resources and characterization facilities for the nanodiamond samples. All scientific content, data interpretation, and final text were critically reviewed and verified by the authors.

**Funding:** This research has received funding from the German Federal Ministry of Research, Technology and Space (project DiamondNanoNMR, No. FKZ13N16297), the European Union and from tax funds on the basis of the budget adopted by the Saxon State Parliament (sub-project of the measure with the application no. AZ 100760036).

**Author contributions:** DTM, BS, and NA conceptualized the study. DTM, BS, ETG, CL, and NA developed the methodology. DTM, BS, and ETG developed software routines. BS, ETG, and MS performed formal analysis. DTM, BS, DW, OG, DAGV, MKC, SPC, CL, SKSR, and NA conducted investigations. CL provided, treated, and characterized the nanodiamond samples. NA provided lab resources, funding, project administration, and overall supervision. DTM, BS, ETG, and NA wrote the original draft. All authors reviewed, edited, and approved the final manuscript.

**Competing interests:** NA, BS, and DTM are inventors on two pending European patent applications (EP26183325.5 and EP26196338.3) filed by Leipzig University related to the technology described in this work.

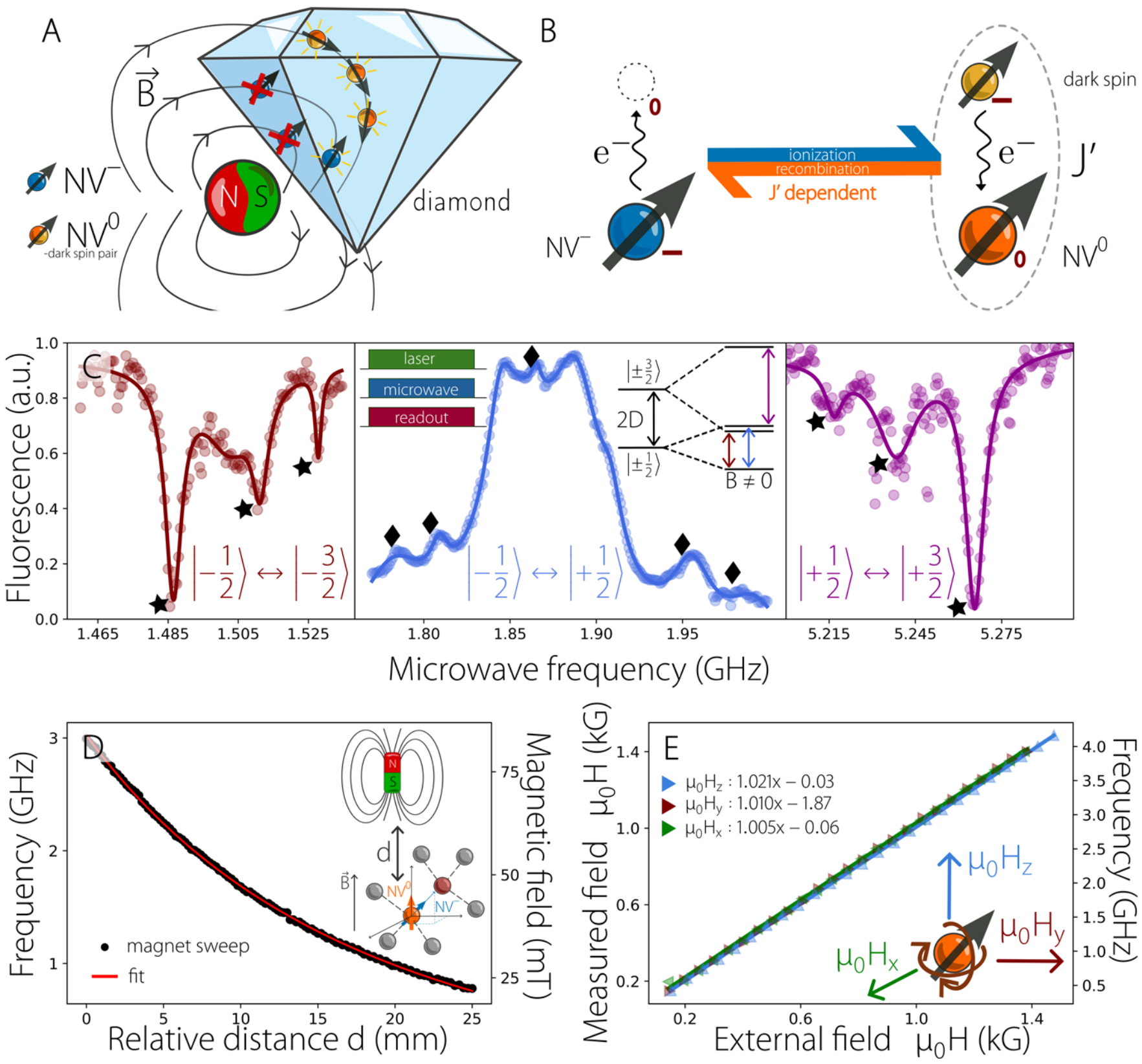

**Fig. 1. Platform and isotropic spin-to-charge (SC-)ODMR. (A)** Schematic comparison of $NV^0$- and $NV^-$-based quantum magnetometry. Unlike the $NV^-$ center, whose quantization axis is fixed to the diamond lattice, the $NV^0$-dark spin pair's axis aligns dynamically with the local magnetic field, yielding an isotropic response via its central $|+1/2\rangle \leftrightarrow |-1/2\rangle$ transition. **(B)** Schematic of SC-ODMR mechanism: Charge cycles form an $NV^0$-dark spin pair with total spin angular momentum $J'$. Charge recombination is allowed for certain $J'$ states but blocked for others, mapping the spin pair's state onto the $NV^-/NV^0$ charge ratio. **(C)** CW SC-ODMR spectra of the 3/2 spin of $NV^0$ and the dark spin at a magnetic field of 667 G with a measured ZFS of $2D \approx 3.370$ GHz. Peaks corresponding to P1 center's (♦) and $NV^0$'s (★) hyperfine coupling to nitrogen nuclear spins are resolved. Insets: Experimental sequence and energy level scheme of spin-3/2 in $NV^0$'s metastable state. **(D)** Scalar magnetometry of a permanent magnet, using the central $|+1/2\rangle \leftrightarrow |-1/2\rangle$ transition. Red curve depicts an inverse-cubic fit. **(E)** Isotropic magnetic response of the $NV^0$-dark spin central transition. Resonance frequency shifts remain invariant under in-plane (x/y) and out-of-plane (z) field orientations, applied with a vector magnet.

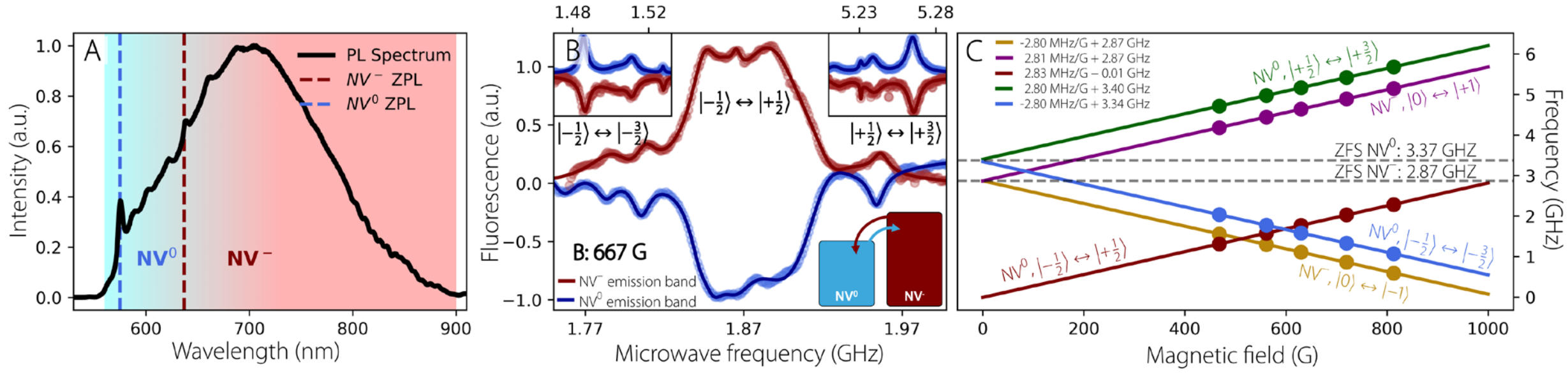

**Fig. 2. Dual-band spin readout and multi-species magnetometry. (A)** Fluorescence emission spectrum displaying distinct zero-phonon lines (ZPLs) and phonon sidebands for $NV^0$ (575 nm) and $NV^-$ (637 nm), enabling simultaneous dual-band detection. **(B)** SC-ODMR spectra of the 3/2 $NV^0$-dark spin system. The central $|+1/2\rangle \leftrightarrow |-1/2\rangle$ transition exhibits positive contrast in the $NV^-$ fluorescence channel and negative contrast in the $NV^0$ channel, whereas the outer $|\pm1/2\rangle \leftrightarrow |\pm3/2\rangle$ transitions show reversed polarity. Matching resonance frequencies across both detection bands confirm a spin-to-charge conversion mechanism. Spectral substructure arises from hyperfine coupling to the intrinsic $^{14}N$ nuclear spin and of proximal substitutional nitrogen (P1) centers (see sections S1-S2). **(C)** Resonance frequencies of $NV^-$ (spin-1) and $NV^0$ (spin-3/2) transitions as a function of applied magnetic field. Linear Zeeman scaling with $g \approx 2$ demonstrates simultaneous multi-species magnetometry from a single optical spot. Extracted $g$-factors are listed in table S2.

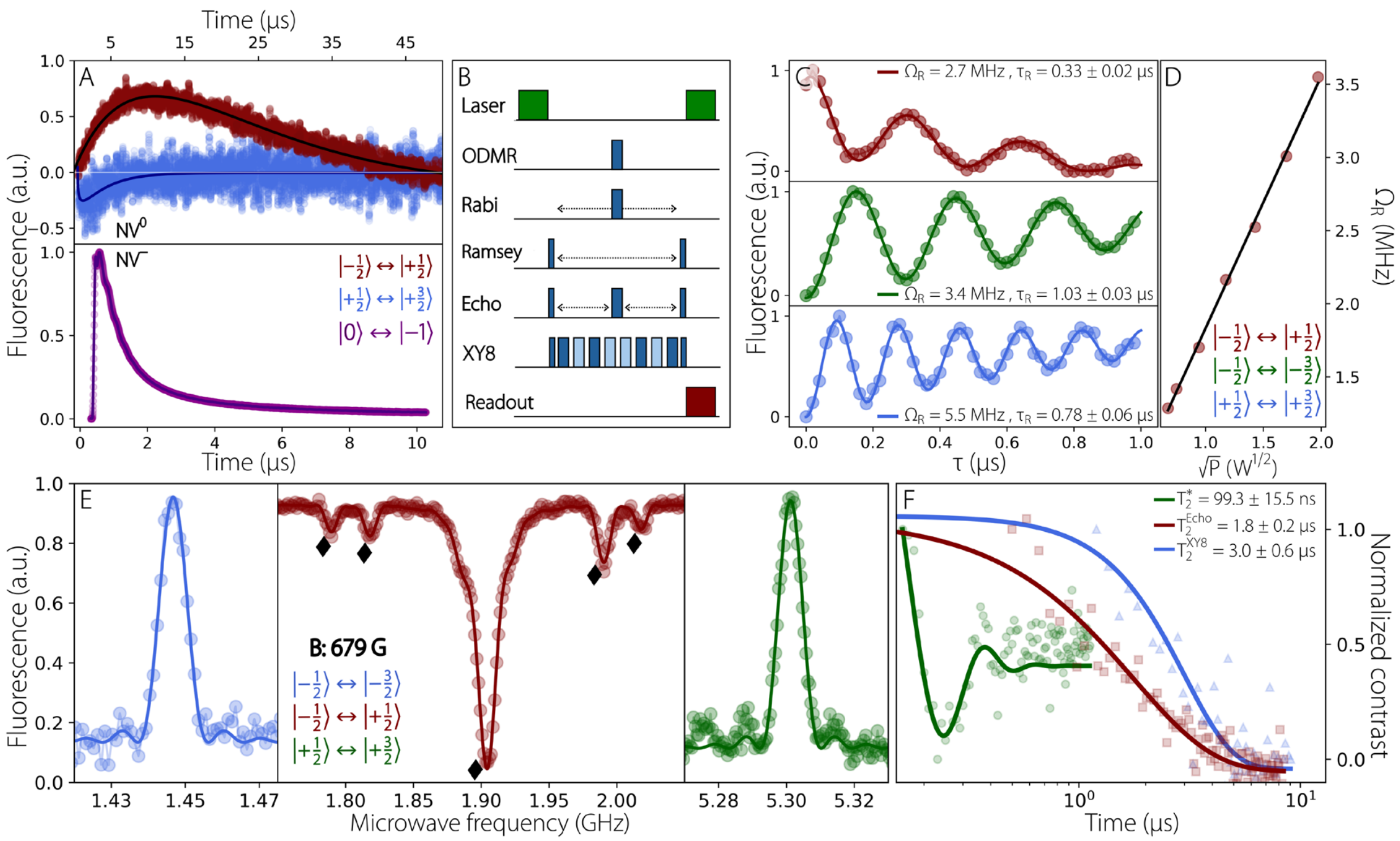

**Fig. 3. Coherent control of the NV⁰ spin 3/2-dark spin pair under ambient conditions. (A)** Decay of fluorescence contrast following microwave excitation resonant with the central (|−1/2⟩ ↔ |+1/2⟩) and outer (|+1/2⟩ ↔ |+3/2⟩) NV⁰ transitions. Signal decay occurs on a 5 µs timescale for the outer transition, whereas signal persists beyond 50 µs for the center isotropic transition. The case for NV⁻ is also shown for comparison. **(B)** Diagram of microwave and optical pulses used for coherent manipulation, including pulsed ODMR, Rabi oscillations, Ramsey interferometry, Hahn echo, and XY8 dynamical decoupling (MW pulses: narrow blue $^{\pi_x}/_2$, blue $\pi_x$; light blue $\pi_y$). **(C)** Rabi oscillations for the central and outer transitions ($\Omega_R$: Rabi frequency; $\tau_R$: Oscillation decay). **(D)** Rabi frequency as a function of microwave power with a linear fit (black line), showing the expected square-root-scaling ($\propto\sqrt{P}$). **(E)** Pulsed ODMR spectra of NV⁰-dark spin system at 679 G, resolving hyperfine splittings characteristic of P1 centers' coupling to their intrinsic nuclear spins (♦). **(F)** Room-temperature spin coherence of the central |−1/2⟩ ↔ |+1/2⟩ transition. Ramsey interferometry, Hahn echo, and XY8-2 dynamical decoupling yield $T_2^* = 99$ ns, $T_2 = 2$ µs, and $T_2 = 3$ µs, respectively.

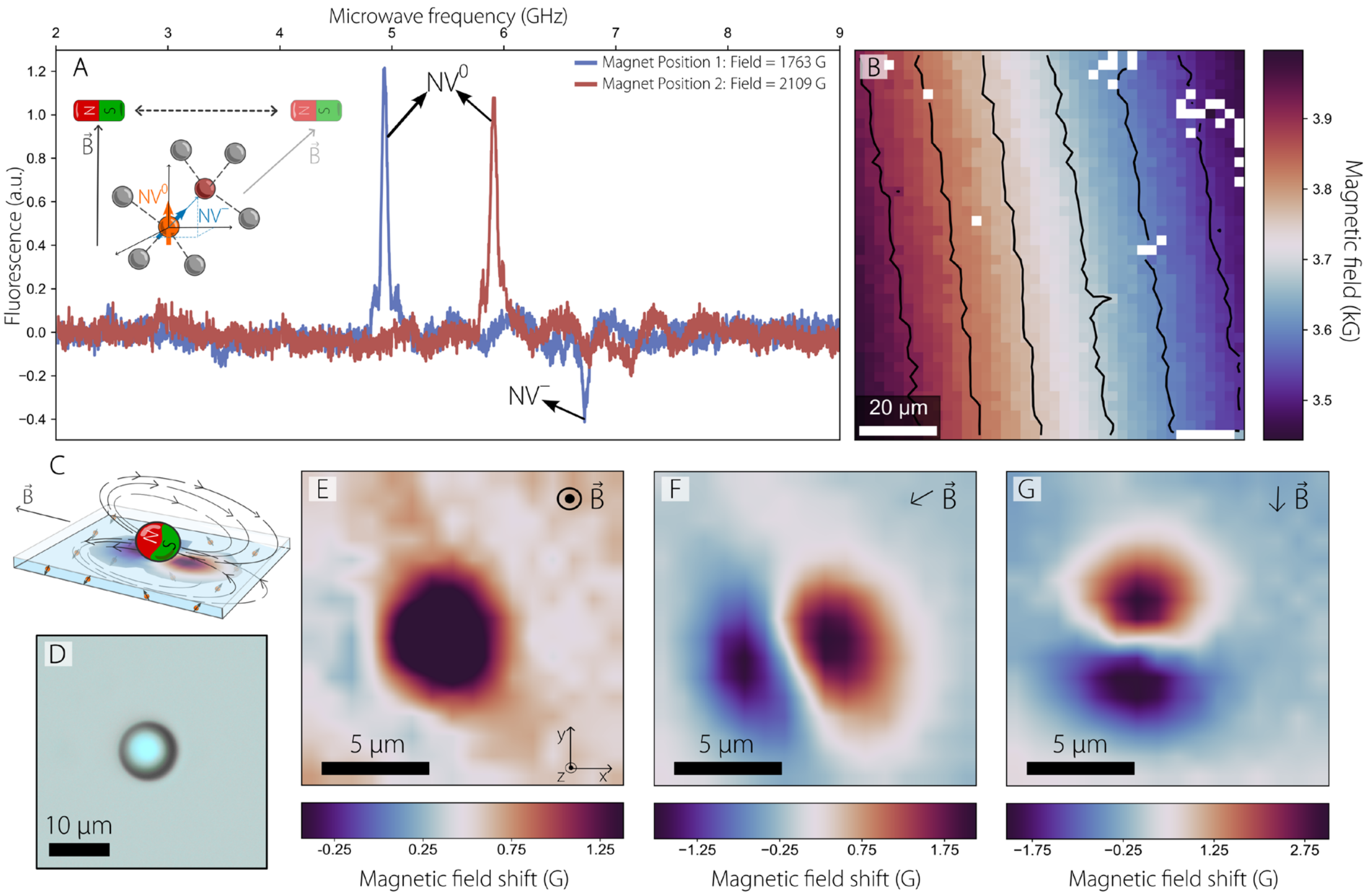

**Fig. 4, Isotropic magnetic field mapping of individual paramagnetic microbeads. (A)** CW ODMR spectra recorded at two locations near a permanent magnet. At ~1760 G, both $NV^-$ and the $NV^0$-dark spin pair yield clear resonance signals. At ~2110 G, the increased off-axis field component induces spin-state mixing in $NV^-$, quenching its signal. The $NV^0$-based sensor maintains a robust resonance. **(B)** Magnetic mapping near a permanent magnet using the $NV^0$-dark spin pair. Scanning a 100 µm × 100 µm area on the diamond yields the local magnetic field magnitude maps, independent of field orientation or gradient strength. **(C)** Illustration of $NV^0$ centers in a diamond substrate probing stray magnetic field lines of an in-plane magnetized paramagnetic microbead. **(D)** Brightfield image of a single paramagnetic microbead (nominal diameter ~10 µm) on the diamond substrate. **(E)** Magnetic field profile under an out-of-plane bias field (≈ 2694 G), displaying a symmetric Gaussian profile. (**F**) and **(G)** Magnetic field profiles under nearly orthogonal in-plane bias fields along the x- (≈ 2793 G) and y-axes (≈ 2760 G), respectively, revealing characteristic dipole field signatures. All three images are produced by scanning a 15 µm × 15 µm area with a 1 µm resolution, with pixels interpolated for improved visualization.

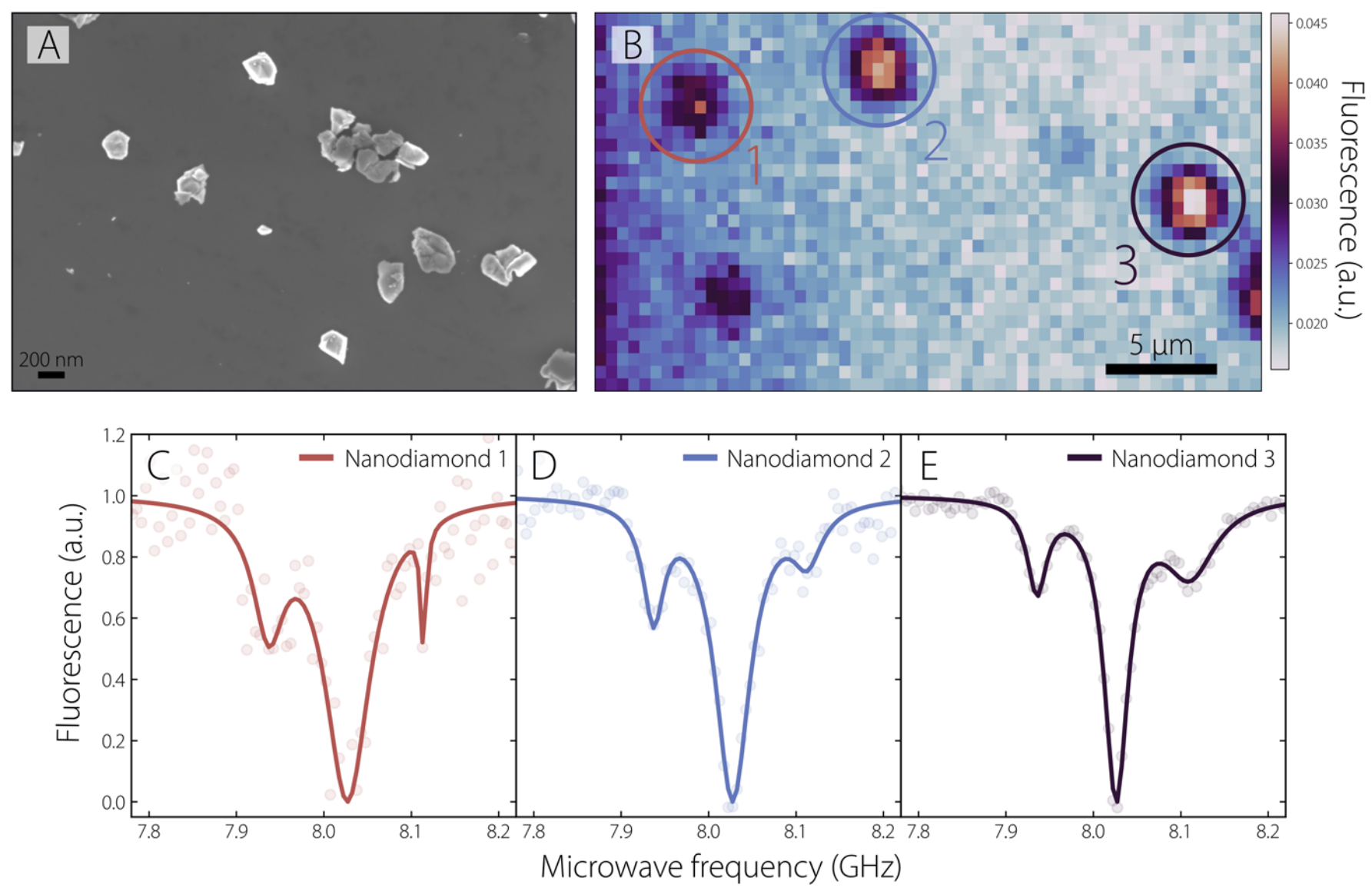

**Fig. 5, Isotropic spin readout of NV⁰-dark spin pairs in nanodiamonds at room temperature. (A)** Scanning electron microscopy (SEM) image showing the topography of nanodiamonds with a nominal size of 250 nm. **(B)** Confocal fluorescence map of multiple individual nanodiamonds, acquired with a longpass filter selective to $NV^-$-fluorescence. **(C to E)** CW SC-ODMR spectra of $NV^0$-dark spin system recorded from three single nanodiamonds marked in (B) under an out-of-plane bias magnetic field of 2865 G using an optical bandpass filter optimized for $NV^0$-fluorescence. Despite the random crystallographic orientations of the individual particles, all three nanodiamonds exhibit identical resonance frequencies and similar linewidths for the central $|+1/2\rangle \leftrightarrow |-1/2\rangle$ transition, confirming alignment-free spin readout. Additional details on nanodiamond preparation and measurement conditions are provided in the Supplementary Materials.